\documentclass [twocolumn,
amsymb,amsmath,nobibnotes,superscriptaddress,aps,prl ]{revtex4}
 \usepackage {bm}
 \usepackage{graphicx}
 \usepackage{amsmath}

\begin{document}
\title{Stacking order, interaction and weak surface magnetism in layered graphene sheets
}

\author{Dong-Hui Xu}
\affiliation{Department of Physics, Zhejiang University, Hangzhou,
310027, China}

\author{Jie Yuan}
\email[Yuan and Xu contributed equally to this work.]{}
\affiliation{Department of Physics, and Center of Theoretical and
Computational Physics, The University of Hong Kong,
 Hong Kong, China}

 \author{Zi-Jian Yao}
\affiliation{Department of Physics, and Center of Theoretical and
Computational Physics, The University of Hong Kong,
 Hong Kong, China}

\author{Yi Zhou}
 \affiliation{Department of Physics, Zhejiang University,  Hangzhou, 310027,  China}

 \author{Jin-Hua Gao}
 \email {jhgao1980@gmail.com}
 \affiliation{Department of Physics, Huazhong University of Science and Technology, Wuhan, China}
 \affiliation{Department of Physics, and Center of Theoretical and Computational Physics, The University of Hong Kong,
 Hong Kong, China}

 \author{Fu-Chun Zhang}
 \email{fuchun@hku.hk}
 \affiliation{Department of Physics, and Center of Theoretical and Computational Physics, The University of Hong Kong,
 Hong Kong, China}
\affiliation{Department of Physics, Zhejiang University,  Hangzhou,
310027, China}

\begin{abstract}

Recent transport experiments have demonstrated that the rhombohedral
stacking trilayer graphene is an insulator with an intrinsic gap of
6meV and the Bernal stacking trilayer one is a metal. We propose a
Hubbard model with a moderate $U$ for layered graphene sheets, and
show that the model well explains the experiments of the stacking
dependent energy gap. The on-site Coulomb repulsion drives the
metallic phase of the non-interacting system to a weak surface
antiferromagnetic insulator for the rhombohedral stacking layers,
but does not alter the metallic phase  for the Bernal stacking
layers.
\end{abstract}

\maketitle
 In the past several years, the rapid development in preparing few layer graphene samples
 has promoted great theoretical~\cite{fzhang,henrard06,min08,rmp,peeters,koshino,mccann,dft,guo1,guo2,chou,dora} and
 experimental~\cite{tarucha,lzhang,kumar,herrero,heinz11,heinz12,jhang,wbao,mak} interests in such
novel quasi-two-dimensional electron systems. The few layer graphene
may be a platform for many new physics issues and is of
potential application in electronics. One peculiar feature of the
layered graphene system is the stacking order, which offers a new route to manipulate
the electronic properties in graphene layers.

The Bernal (or ABA) stacking and the rhombohedral (or ABC) stacking
are two stable stacking orders observed in experiments. As shown in
Fig. 1, in either ABA or ABC stacking order, the second graphene
sheet is shifted by one bond length along the C-C bond direction.
The third graphene sheet is shifted back and aligned with the first sheet in the ABA stacking, while is
shifted further by one more bond length along the same direction in the
rhombohedral stacking. So the ABA stacking order is ABABAB$\cdots$,
and the rhombohedral stacking is ABCABC$\cdots$. The trilayer
graphene system is the minimal structure relevant to the stacking
orders.

The electronic structures of the graphene layers strongly depend on
their stacking orders~\cite{rmp,henrard06,min08}. In the ABA
stacking $N$-layer system, there are $N/2$ electron-like and $N/2$
hole-like parabolic sub-bands touching at $\epsilon=0$ for even $N$,
and an additional sub-band with a linear dispersion for odd $N$. The
states in all the sub-bands are bulk states extended to all the
layers. In the ABC stacking layers, the low energy electronic
structure is described by two sub-bands with dispersion $\epsilon
\sim k^N$ near the points $K$ and $K'$ in the 2D Brillouin zone.
These low energy states are localized on the outermost layers, and
are zero modes on the surfaces protected by the
topology~\cite{flat,sc}. In two dimension, the dispersion of
$\epsilon ( k) \sim k^N$ gives a density of states $D(\epsilon) \sim
\epsilon^{-1 + 2/N}$, which is divergent for $N \geq 3$ at
$\epsilon=0$. This indicates a strong instability toward symmetry
broken states\cite{fzhang,sc}.

\begin{figure}[hptb]
\centering
\includegraphics[width=8.5cm]{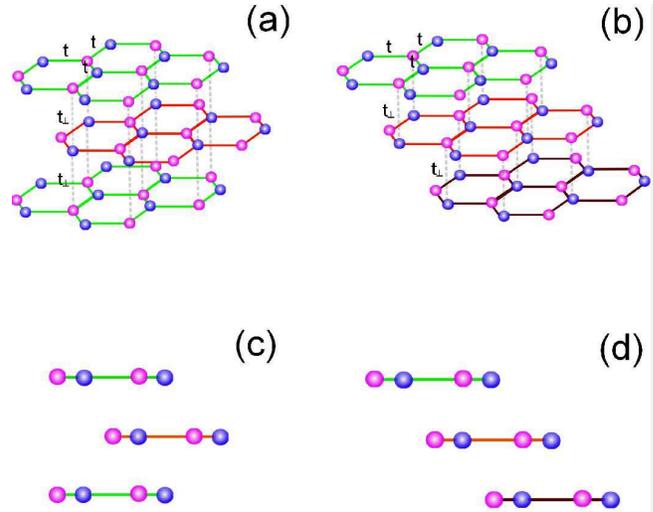}%
\caption{(Color online).  Schematic diagrams of trilayer graphene
sheets. (a): Bernal (ABA) stacking and  (b): rhombohedral (ABC)
stacking.  (c) and (d) are their side views. Blue and pink colors
represent carbon atoms on sub-lattices $A$ and $B$, respectively.}
\label{fig1}
\end  {figure}

Trilayer graphene systems are of particular interest for it
represents the simplest case for the stacking dependent graphene.
Very recently, a stacking dependent intrinsic gap in trilayer
graphene has been observed in the transport
measurement~\cite{wbao,jhang}. In the charge neutral case, namely
undoped trilayer samples,
the experiments indicate that the
ABA stacking trilayer graphene  is metallic, whereas the ABC
stacking trilayer graphene is insulating with an intrinsic gap about
6 $\textrm{meV}$. Since the non-interacting electronic structure of
both 
stacking orders are gapless hence metallic, the experimental
observation of the gap in 
ABC stacking trilayer is in sharp contraction with the
non-interaction picture and points to the importance of the
interaction in these systems.

In this Letter, we propose that the observed stacking-dependent
metallic or insulting states can be explained by a Hubbard model with a
moderate on-site Coulomb repulsion $U$. We use a
self-consistent mean field theory to show that the ground state of
the ABC stacking trilayer is a weak anti-ferromagnet with
opposite ferrimagnetic orderings on the top and bottom layers, due to the
divergent density of states in the metallic phase. The magnetic
ordering opens a gap, which is in good comparison with the experimental data.
Our theory shows that the metallic phase of the ABA stacking
trilayer is stable against a moderate Hubbard $U$ due to the
non-divergent density of state. Our theory is extended to study
stacking-dependent graphene systems for larger numbers of layers. We
have found that it is a general property for the ABC stacking
graphene layers that the on-site Hubbard $U$ opens a gap at the Fermi
level and leads to a weak surface antiferromagnetic state. Our
results can be further tested in future experiments.

We model $N$ layer graphene systems by using a Hubbard model $H=H_0
+ H_U$, where $H_0 = H_{\textrm{intra}}+H_{\textrm{inter}}$ is a
tight binding Hamiltonian to describe the kinetic term of the system
and $H_U$ describes the on-site Coulomb repulsion. The chemical
potential is set to zero, and the average electron per site is one.
 The intralayer hopping term $H_{\textrm{intra}}$ is the tight-binding
Hamiltonian of independent graphene sheets. For simplicity, we only
include nearest neighbor hoppings~\cite{footnote}
\begin{equation}
 H_{ \textrm{intra} }= -t \sum_{l \langle ij \rangle \sigma }
 \{ a^\dagger_{l \sigma}(i)b_{l \sigma}(j) + h.c \}
\end{equation}
where $a_{l\sigma}(i)$ and $b_{l\sigma}(j)$ are the annihilation
operators of an electron on sublattices $A$ and $B$, respectively.
$l$ denotes the layer index running from 1 to $N$, and $\langle ij \rangle$ nearest neighbor pairs, and $\sigma$ the spin.
 $H_{\textrm{inter}}$ describes the interlayer
hopping given by
\begin{equation}
H^\textrm{R,B}_{\textrm{inter}}=t_\perp \sum_{\langle ll' \rangle,
\langle ii' \rangle \sigma} \{
a^\dagger_{l\sigma}(i)b_{l'\sigma}(i') + h.c \}.
\end{equation}
for the rhombohedral or Bernal stacking orders. Here, $\langle ll'
\rangle$ is summed over the two adjacent layers, and
 $\langle ii' \rangle$ is summed over two sites aligned in adjacent
layers as shown in Fig. 1. The Hubbard term $H_U=
U\sum_{li}n_{l\uparrow}(i)n_{l\downarrow}(i)$ will be approximated
by a mean field Hamiltonian,
\begin{equation}
H^{\textrm{MF}}_U=U\sum_{l ,i \sigma} \langle n_{l\sigma}(i)\rangle
n_{l\bar{\sigma}}(i),
\end{equation}
where $\bar{\sigma} = -\sigma$. $\langle n_{l\sigma}(i) \rangle$ is determined
self-consistently. We consider a spin density wave state and introduce
two mean fields
 on each layer $l$, one for sublattice $A$ and one for sublattice $B$, $\langle n_{l\uparrow}^{A,B} \rangle$.
 The  mean fields for spin down are related to the spin-up ones,
 $\langle n_{l\downarrow}^{A,B} \rangle = 1 - \langle n_{l\uparrow}^{A,B} \rangle$.
  Note that we have examined possible
 charge density wave states within the model and found no evidence for that.

 \begin{figure}[hptb]
\centering
\includegraphics[width=8.5cm]{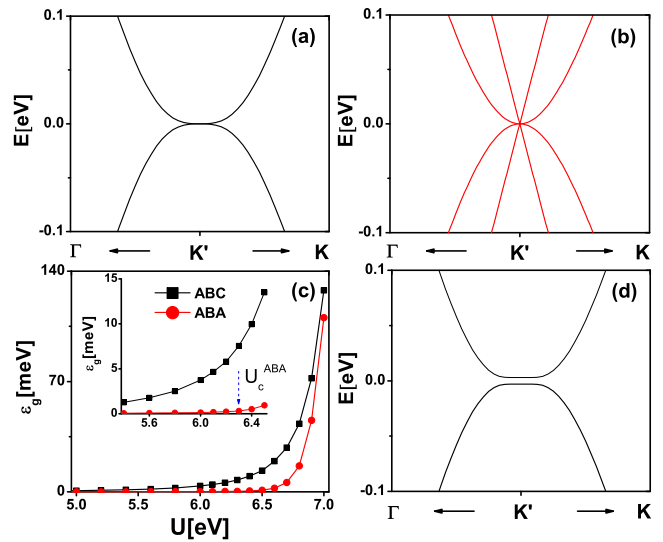}
\caption{(Color online). Low energy electronic bands in ABC (panel
a) and ABA stacking (panel b) trilayer graphene at  $U=0$. Panel (c): mean field energy gaps as functions of U for ABC and ABA
trilayer graphene. Panel
(d): energy bands of ABC stacking with $U=6.2\textrm{eV}$.
 The parameters are $t =3.16\textrm{eV}$ and
$t_\perp =0.39\textrm{eV}$.} \label{band}
\end {figure}

We first examine the trilayer graphenes ($N=3$). The energy bands
for the non-interacting models are shown in Fig. 2 (a) for ABC
stacking and in Fig. (b) for ABA stacking with parameters $t=3.16$eV
and $t_{\perp}=0.39$eV. The dispersion in (a) is $\epsilon \sim k^3$
at small $k$ for both conduction and valence bands, which gives rise
to a divergent density of states $D(\epsilon)= \epsilon^{-1/3}$ at
$\epsilon =0$, and the wave functions for $k$ near the $K$ or $K'$
points are localized on the outer surfaces. The energy bands in (b)
consist of a parabolic and a linear dispersions, both of which are
not localized on the outer surfaces, and the density of states is a
constant at $\epsilon =0$. In the presence of the Hubbard $U$, the
spin density wave ordering occurs at any $U>0$ for the ABC stacking
case, and only at $U > U_c^{\textrm{ABA}} \approx 6.3$eV. The energy
gaps associated with the spin density wave orderings are plotted in
Fig. 2(c) as functions of $U$ for both the ABC and ABA stacking
orders. At $U < U_c^{\textrm{ABA}}$, the energy bands of the ABA
stacking trilayer graphene are rigid against the Hubbard $U$. The
sharp distinction between the ABC and ABA stacking trilayer graphene
is attributed to their different density of states near the Fermi
level. The divergent density of states of the surface zero modes for
the ABC stacking graphene, protected by the momentum topology,
actually induces its sensitivity to the interaction.

More quantitatively, there are three distinguished regions in $U$ for the gaps.
At $U<5.5\textrm{eV}$, the energy gap is zero for the ABA stacking and is tiny for the ABC stacking.
At $5.5\textrm{eV} < U < 6.4\textrm{eV}$, the gap size grows rapidly to be observable (several meV) for the ABC stacking,
but remains zero or tiny for the ABA stacking. In this region, the ABC stacking trilayer is insulating with an
observable gap while the ABA stacking trilayer remains conduct.
At $ U > 6.4\textrm{eV}$, the gaps for both ABC and ABA stacking
orders become observable, and become insulating. Actually the gaps
for the two stacking orders become similar at $U > 7\textrm{eV}$ as
we can see from Fig. 2(c). Experimentally, the transport data shows
ABC stacking trilayer graphene is an insulator with a gap of 6meV
and the ABA stacking trilayer is metallic. In comparison with the
experiments, the mean field calculations of the Hubbard model
suggest that the Hubbard $U$ is within the interval of a moderate
values $5.5\textrm{eV}<U<6.4\textrm{eV}$, i.e.
$1.74t<U<2.03t$.

In Fig. 2(d), we show the calculated quasi-particle dispersion for
the ABC stacking trilayer graphene for a choice of $U=6.2$eV. The
corresponding gap is estimated to be $\epsilon_{g} \approx 5.8$meV.
Our model and the calculations well explain the recent experiments
showing the stacking-dependent energy gap in trilayer graphene. The
experimentally observed energy gap may be used to estimate the value
of $U$. Our mean field theory suggests that $U \approx 6.2$eV. More
accurate numerical simulation may improve this estimate.

\begin{figure}[hptb]
\centering
\includegraphics[width=7.5cm]{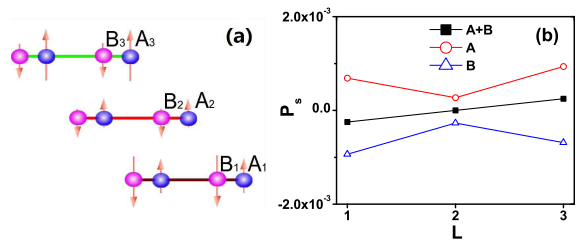}
\caption{(Color online). Schematic illustration of the spin
orderings (panel a) and calculated net spin polarization per site in
each layer $L$ (panel b) in the charge neutral ABC stacking trilayer
graphene.  In the calculations, $U=6.2\textrm{eV}$, and the hopping
parameters are the same as in Fig. 2.}
\end{figure}

We now discuss the spin density state and the spin polarization of
the ABC stacking layer. From the self-consistent mean field theory
we obtain the site spin polarization on sublattice $A$ or $B$,
defined as $P_s(l, i) =\langle n_{l\uparrow}^{A,B} \rangle
- \langle n_{l\downarrow}^{A,B} \rangle$. The calculated spin
polarizations are plotted in Fig. 3(b), and the spin structure in
the trilayer graphene is schematically illustrated in Fig. 3(a). The
spin ordering is antiferromagnetic, where the neighboring spins
(intra- or inter- layer) are anti-parallel to each other. However,
there is a net spin polarization on the top or bottom layer, so each
surface shows ferrimagnetic ordering. The spin polarization is
mainly distributed on the two outer surfaces. In each layer, the
spin polarizations on sublattices A and B have opposite directions.
The net spin polarization is zero in the mid layer, and has opposite
sign in the top or bottom layer. There is a symmetry of combined
inversion and time reversal: $P_s(l=1, i\in A (B))=
P_s(l=3, i\in B (A))$. Note that the average spin
polarization of the whole system is zero. For the parameters given
in Fig. 3(b), the site spin polarizations in the top layer are about
$6.9\times10^{-4}$ and $-9.6\times10^{-4}$ on sublattices $A$ and
$B$, respectively, and the net spin polarization is $-2.5\times
10^{-4}$ per site in average, which gives a surface magnetization
$0.005 \mu_B / \textrm{nm}^2$. The weak surface magnetization on
the ABC trilayer graphene is in analogy with the ferromagnetic edge
states in graphene zigzag ribbon~\cite{son}, in which the density of
states of the flat band edge states is divergent, inducing the edge
spin polarization in the presence of a weak interaction. The
interaction induced
gap in graphene zigzag ribbon has been confirmed in a
recent STM experiment~\cite{dai}.

\begin{figure}[hptb]
\centering
\includegraphics[width=8.5cm]{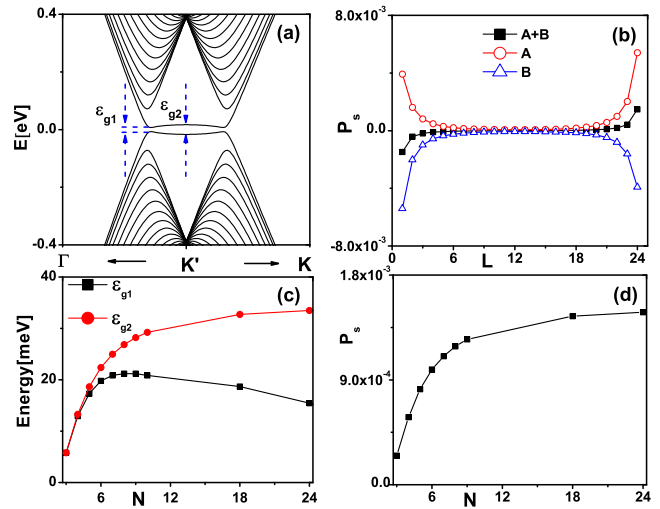}%
\caption{(Color online). ABC stacking $N$-layer graphene.  Energy
band (panel a) and spin polarization (panel b) obtained in mean
field theory for $N=24$. The gaps $\epsilon_{g1}$ and
$\epsilon_{g2}$ (panel c) and spin polarization (panel d) as
functions of $N$. $U=6.2eV$, and the hopping parameters are the same
as in Fig. 2. } \label{spin}
\end  {figure}

We now discuss $N > 3 $ graphene layers. For the ABC stacking
graphene layers and in the charge neutral case, there is always an
interaction ($U$) induced gap at the Fermi level with spontaneous
surface spin density wave ordering. In Fig. 4(a) and (b), we show
the results of $N= 24$ layers as an example. Since the minimum band
gap ($\epsilon_{g1}$) is no longer at the $K$ or $K^\prime$ points
as we can see from Fig. 4(a), we introduce the second gap
$\epsilon_{g2}$ for the energy gap at the $K$ or $K^\prime$ points.
As shown in Fig. 4(b), the spin polarization is localized near the
surfaces. In Fig. 4 (c), we present the energy gaps as functions of
$N$. As $N$ increases, $\epsilon_{g1}$ first increases to approach
its maximum about $20\textrm{meV}$ at $N=9$, then decreases to a
value of $15 \textrm{meV}$ at $N= 24$. On the other hand,
$\epsilon_{g2}$ increases with the layer thickness $N$ and reaches a
saturated value about $33\textrm{meV}$. The $N$-dependent spin
polarization is shown in Fig. 4 (d), which increases with the layer
thickness, and approaches to a saturated value, which is at least 5
to 6 times of the surface magnetization in the trilayer case. Note
that the first principle calculations involving a local spin density
approximation has been applied to study 8 layers ABC stacking
graphene \cite{dft}, and reported a spin density wave ground state.
Their result is consistent with the results of the Hubbard model
proposed here, while our results are more general and distinguishes
different stacking orders.

We have also applied the mean field theory to study $N$-layer
graphene of the ABA stacking. The results are similar to the
trilayer case and there is a threshold $U_c^{\textrm{ABA}}$, which is weakly
$N$ dependent and remains to be finite at large $N$, the metallic
phase is stable against the spin density wave ordering at $U <
U_c^{\textrm{ABA}}$.

We argue that the mean field theory should give a qualitatively or
semi-quantitatively correct physics on the stacking dependent
instability, or the insulating or metallic states in layered
graphene, while more accurate  calculations may refine the estimate
of the value of $U$. We remark that the proposed Hubbard model
with a moderate $U$ should capture the most important physics for the
stacking dependent ground states in layered graphene. The intersite
Coulomb repulsion has tendency to drive the metallic phase to a
charge density wave state, which is not compatible with the on-site
$U$ studied in the present work. Since the intersite repulsion
 is relatively weaker than the on-site Coulomb repulsion $U$, we may argue that that term may not be relevant.
More exotic states such as quantum spin Hall state and anomalous
Hall state have been proposed in models with spin-orbit coupling or
intersite interaction on honeycomb lattice~\cite{sczhang,kane}. The
possible realization of these exotic phases in layered graphene will
be highly interesting. In view of the very weak spin-orbit coupling
in graphene\cite{so}, more detailed study will be needed to explore
the possibility.

In summary, we have proposed a Hubbard model with a moderate $U$ to
describe N layer graphene, and applied a mean field theory to study
the ground state and the excited energy gap of the charge neutral
systems. The ground state of the non-interacting graphene layers
are metallic, whose density of states at the Fermi level is
divergent as $D(\epsilon) \sim \epsilon^{-1+2/N}$ for rhombohedral
(ABC) stacking, and is a constant for Bernal (ABA) stacking. The
metallic ground state of the ABA stacking layer is stable against
the on-site Coulomb repulsion for moderate value of $U$ below a
threshold $U_c^{\textrm{ABA}}$ about 6-7 eV. The metallic state of the ABC
stacking layer is found to be unstable against any repulsion $U$ due
to the divergent density of states at zero energy. Its ground state
is surface antiferromagnetic state with opposite ferrimagnetism on
top or bottom surfaces, which opens a gap. The energy gap is
estimated to be $5.8$meV for $U=6.2$eV for $N=3$. Our model and
calculations well explain the recent transport experiments, showing
that the ABC stacking trilayer graphene is an insulator with a gap
about $6$ meV, and ABA stacking trilayer graphene remains to be
metallic. The spin polarization in the spin ordered state is found
to be weak, but should be measurable.

We acknowledge part of financial support from HKSAR RGC grant HKU
701010 and CRF HKU 707010. DHX and YZ are supported by National
Basic Research Program of China (973 Program, No.2011CBA00103), NSFC
(No.11074218) and the Fundamental Research Funds for the Central
Universities in China.


\begin{thebibliography}{}
\bibitem{footnote}
The small values of longer range hopping integrals do not change the basic conclusion presented here.
\end{thebibliography}
\end{document}